\documentstyle[12pt,psfig,epsf]{article}
\newcommand{\be}{\begin{equation}}
\newcommand{\ee}{\end{equation}}

\newcommand{\beqa}{\begin{eqnarray}}
\newcommand{\eeqa}{\end{eqnarray}}
\newcommand{\nn}{\nonumber}

\newcommand{\eqref}[1]{(\ref{#1})}

\def\boxit#1{\vbox{\hrule\hbox{\vrule\kern8pt
\vbox{\hbox{\kern8pt}\hbox{\vbox{#1}}\hbox{\kern8pt}}
\kern8pt\vrule}\hrule}}
\def\mathboxit#1{\vbox{\hrule\hbox{\vrule\kern8pt\vbox{\kern8pt
\hbox{$\displaystyle #1$}\kern8pt}\kern8pt\vrule}\hrule}}

\def\IB{\relax\hbox{$\inbar\kern-.3em{\rm B}$}}
\def\IC{\relax\hbox{$\inbar\kern-.3em{\rm C}$}}
\def\ID{\relax\hbox{$\inbar\kern-.3em{\rm D}$}}
\def\IE{\relax\hbox{$\inbar\kern-.3em{\rm E}$}}
\def\IF{\relax\hbox{$\inbar\kern-.3em{\rm F}$}}
\def\IG{\relax\hbox{$\inbar\kern-.3em{\rm G}$}}
\def\IGa{\relax\hbox{${\rm I}\kern-.18em\Gamma$}}
\def\IH{\relax{\rm I\kern-.18em H}}
\def\IK{\relax{\rm I\kern-.18em K}}
\def\IL{\relax{\rm I\kern-.18em L}}
\def\IP{\relax{\rm I\kern-.18em P}}
\def\IR{\relax{\rm I\kern-.18em R}}
\def\IZ{\relax\ifmmode\mathchoice
{\hbox{\cmss Z\kern-.4em Z}}{\hbox{\cmss Z\kern-.4em Z}}
{\lower.9pt\hbox{\cmsss Z\kern-.4em Z}} {\lower1.2pt\hbox{\cmsss
Z\kern-.4em Z}}\else{\cmss Z\kern-.4em Z}\fi}

\def\II{\relax{\rm I\kern-.18em I}}

\def\CE {{\cal E}}

\def\CO {{\cal O}}

\def\CS {{\cal S}}

\def\p{\partial}

\begin{document}

\hfill  NRCPS-HE-36-2010

\vspace{5cm}
\begin{center}
{\Large \it Large Spin Behavior of Anomalous Dimensions\\

\vspace{1cm}

 and \\

\vspace{1cm}
 Short-long Strings Duality
}


\vspace{2cm}

{\sl George Georgiou and George Savvidy\\
Demokritos National Research Center\\
Institute of Nuclear Physics\\
Ag. Paraskevi, GR-15310 Athens,Greece \\
\centerline{\footnotesize\it E-mail: georgiou@inp.demokritos.gr, savvidy@inp.demokritos.gr}
}
\end{center}
\vspace{60pt}

\centerline{{\bf Abstract}}

We are considering the semi-classical string soliton solution of Gubser, Klebanov
and Polyakov
which represents highly excited states on the leading Regge trajectory, with large spin in $AdS_5$.
A prescription relates this soliton solution with the
corresponding field theory operators with many covariant derivatives,
whose anomalous scaling dimension grows logarithmically with  the space-time spin. We
develop an iteration procedure which, in principle, allows to derive all terms
in the large spin expansion  of the anomalous
scaling dimension of twist two operators at strong coupling.
We explicitly derive the dependence of anomalous
dimension on spin for all leading and next-to-leading orders.
Our string theory results are consistent with the conjectured "reciprocity" relation,
which has been verified to hold in perturbation theory
up to five loops in $N=4$ SYM. We also derive a duality relation between
long and short strings.

\vspace{12pt}

\noindent


\newpage

\pagestyle{plain}
\section{\it Introduction }

It was found in \cite{Gubser:2002tv} that the type IIB string theory soliton solution with large spin on
$AdS_5 \times S^5$
describes the gauge theory operators with many covariant derivatives,
whose anomalous scaling dimension grows logarithmically with the Lorentz spin S.
Operators with many covariant derivatives are present in QCD where they were studied
in the context of deep inelastic scattering and have the following general structure
\cite{Gross:1974cs,Georgi:1951sr,Floratos:1977au}
$$
Tr\{\Phi \nabla_{\lambda_1}...\nabla_{\lambda_S} \Phi\}.
$$
In free field theory such operators have dimension $\Delta=S+2$, while their
space-time spin is $S$.
In the interacting gauge field theories the scaling is violated and the anomalous dimension is
different from zero,
\be
\gamma(\lambda, S) =\Delta- (S+2)= f(\lambda) \ln S + f_0(\lambda) +
f_{11}(\lambda){ \ln   S   \over   S  }
+ f_1(\lambda) {1 \over  S} +...
\ee
Anomalous dimension of these high spin operators is a function of the coupling constant $\lambda^2=  g^2_{YM} N  / 4\pi  $
and of the spin S and  can be computed order by order in the perturbation theory.
The weak coupling corrections are dominated by the leading logarithmic term
$f(\lambda) \ln S$, where $f(\lambda)$ is the so called "cusp anomalous dimension".

The AdS/CFT correspondence \cite{Maldacena:1997re,Gubser:1998bc,Witten:1998qj} allows the strong coupling description
of the same physical quantity, identifying the energy of the string with the
conformal dimension $E=\Delta$ of the dual operator.
Thus the dual, string theory description, provides the strong coupling
behavior of the anomalous dimensions and it appears also to grow logarithmically with spin
\cite{Gubser:2002tv,Alday:2007mf,Frolov:2002av}.
In this article we are intersected in the behavior of subleading in $S$ terms of the
anomalous dimension $\gamma(\lambda, S)$ in the weak and strong coupling regimes.

We have found that the anomalous dimension in the strong coupling regime is given
by the following expression
\be\label{anomalousdimension}
\gamma(\lambda,S) = f \ln  (S/\sqrt{\lambda}) + f_0
+  \sum^{\infty}_{n=1} f_{nn}{ \ln^n (S/\sqrt{\lambda}) \over   S^n }
+  \sum^{\infty}_{n=1} f_{nn-1}{ \ln^n   (S/\sqrt{\lambda})  \over   S^{n+1}  }
+ f_1 {1 \over  S} + f_2 {1 \over  S^2}+...
\ee
where
\beqa\label{coefficiens}
f&=&{\sqrt{\lambda} \over \pi}\nn\\
f_0 &=& {\sqrt{\lambda} \over \pi}~ (2 \rho_0 +  \ln (\pi /2))=
{\sqrt{\lambda} \over \pi}  ~(\ln 8 \pi -1)
\nn\\
f_1 &=& {4 \lambda \over \pi^2} ~(\rho_1  + ~\rho_{11} \ln (\pi/2) )
= { \lambda  \over 2\pi^2}  ~(\ln 8 \pi -1)
\nn\\
f_2 &=& { \lambda^{3/2}  \over 8 \pi^3 }  (- \ln^2 8\pi +{9 \over 2} \ln 8\pi -4\ln 2-1 )
\nn\\
f_{nn} &=&(-1)^{n+1}
{\lambda^{{ n+1\over 2}} \over (2\pi)^{n+1}}{ 2  \over n   }=
\frac{(-1)^{n+1}}{2^n n}~f^{n+1} ,~~~~n=1,2,...
\nn\\
f_{nn-1} &=&  (-1)^{n+1}
{\lambda^{{ n+2 \over 2}} \over (2\pi)^{n+2}}~
(~ {n+4 \over 2} + 2 \sum^{n}_{k=1}{1\over k} -  2 \ln 8\pi  ~),~~~~n=1,2,...
\eeqa
for
\be
\sqrt{\lambda}  \gg 1 ,~~~~ {S \over \sqrt{\lambda}} \gg 1 .
\ee
Two comments are in order. Firstly, notice the appearance of the harmonic sum in the coefficients
of the subleading logarithms $f_{nn-1}$ \footnote{Our $f_{nn-1}$'s are related to the
ones appearing in \cite{Beccaria:2008tg}
by  $f_{nn-1}^{(ours)}=f_{n+1n}^{(\cite{Beccaria:2008tg})}$.}.
Secondly, we should mention that the structure of the large spin expansion at strong
coupling \eqref{anomalousdimension} is the same
with the large spin expansion in perturbation theory \cite{Beccaria:2008tg}.

Using the so-called BES equation \cite{Beisert:2006ez} the cusp anomalous dimension $f(\lambda)$ was thoroughly
studied both in the weak \cite{Beisert:2006ez,Gromov:2009tv} and strong coupling regime \cite{Benna:2006nd,Alday:2007qf,Gromov:2009tv}.
Furthermore, all contributions to the
exact in $S$ anomalous dimension \eqref{anomalousdimension} which are free
from wrapping effects can, in principle, be calculated \cite{Fioravanti:2009xt,Fioravanti:2010ge} by using a linear integral equation previously derived in
\cite{Bombardelli:2008ah} \footnote{ When the length of the operator is $L>3$ there are additional terms in the large $S$ expansion of the anomalous dimension
which scale like $1/\ln^n S$ \cite{Fioravanti:2009ei}.}. Finally, the solution of the thermodynamic Bethe Ansatz (TBA) of \cite{Bombardelli:2009ns} can, in principle,
give the exact result including wrapping effects.
Since our strong coupling coefficients \eqref{coefficiens} include all effects due to wrapping interactions
it would be interesting to solve the TBA equations at strong coupling and compare with our results.

The formulae \eqref{coefficiens} allow to check the reciprocity  relation which assumes the following
functional relation for $\gamma$
\beqa\label{functionalg1}
\gamma(S)=P(S+\frac{1}{2}\gamma(S)),
\eeqa
where $P(S)$ satisfies a ''parity preserving'' or ''reciprocity'' relation \cite{Moch:2004pa,Vogt:2004mw}
which can be cast in the form \cite{Basso:2006nk}
\beqa\label{reciprocity1}
P(S)=\sum_{k=0}^{\infty}\frac{c_k(\ln C)}{C^{2k}}.
\eeqa
In \eqref{reciprocity1}, $C$ is the "bare" quadratic Casimir operator of
the $SL(2,R)$ group given by  $C=S(S+1)$.
The reciprocity relation \eqref{reciprocity1} has been verified to hold in perturbation theory up to three loops in QCD
\cite{Basso:2006nk} and up
to five loops in $N=4$ SYM \cite{Kotikov:2004er,Dokshitzer:2006nm,Kotikov:2007cy,Beccaria:2007bb,Beccaria:2009vt,Lukowski:2009ce}.
It was also found to hold in string perturbation theory at the classical level in \cite{Beccaria:2008tg}
and at one-loop up to order $\frac{1}{S^3}$ in \cite{Beccaria:2008tg,Beccaria:2010ry}.

Equivalently, the first two relations following from \eqref{reciprocity1}
\beqa\label{recTseytlin1}
f_{1}&=& \frac{1}{2}f (f_0 +1) \nonumber \\
f_{21}& =& \frac{1}{16}f [f^3 - 2f^2(f_0 +1) - 16f_{10}]
\eeqa
were verified \cite{Beccaria:2008tg,Beccaria:2009yt}.
Since now we have the infinite series of the coefficients of the subleading
logs in the strong coupling expansion \eqref{anomalousdimension}
it is, in principle, possible to check if the relations following from the reciprocity
property of the function $P$ hold at any level.
By isolating the appropriate terms we get the next relation in the
sequence \eqref{recTseytlin1}. This reads
\beqa\label{recus1}
f_{43}+[f^2_{22}+2f_{11}f_{33}+2 f f_{32}+ f_{44}(2 f_0-\frac{5}{2}f)] +f[3f_{11}f_{22}+\frac{3}{2}f f_{21}+f_{33}(-\frac{35}{8}f+ 3f_0)]
\nonumber \\
+f^2[\frac{3}{4}f^2_{11}+\frac{1}{2}f f_{10}+f_{22}
(-\frac{65}{24}f+\frac{3}{2}f_0)]+f^3[\frac{1}{16}f f_1+f_{11}
(-\frac{125}{12 \,\,16}f+\frac{1}{4}f_0)]~~~~~~~~~~
\nonumber \\
+f^5[\frac{25}{12 \,\,32}f-\frac{1}{32}f_0] =0.~~~~~~~~~~~~~~~~~~~~~~~~~~~~~~~~~~~~~~~~~~~
\eeqa
Now we can use \eqref{coefficiens} to find and substitute the values for the various
coefficients needed in \eqref{recus1} to find that
this equation is, indeed, satisfied.
Thus, we have seen that the constraints coming from the
reciprocity relation \eqref{reciprocity1} are consistent with the infinite series of coefficients
of \eqref{coefficiens}.

In the next section we shall derive a duality relation for
anomalous dimensions. This functional equation defines a map between
$\gamma=\gamma(\lambda,S)$  and $\gamma^{'}= \gamma(\lambda,S^{'})$
at complementary $S, S^{'}$ values of spins. These values of spins are related by:
$$
\frac{S}{\sqrt{\lambda}}   \frac{S^{'}}{\sqrt{\lambda}}\approx \frac{1}{\pi}.
$$
Thus the strings having spins $S$ and $S'$ are complementary-dual to  each other:
$$
 S \gg  \sqrt{\lambda}  \gg 1   ~~~~~~~~~~~\Leftrightarrow ~~~~~~~~~~~~~~~  \sqrt{\lambda}  \gg S   \gg 1 ,
$$
that is in regions where the spin is much larger and much smaller than the coupling constant.

\section{\it Short-long strings duality}

An approximate description of the closed strings on the leading Regge trajectory
is given by folded closed string which spins as a rigid rod around its center
in the warped $AdS_5$ space background with the global metric
\be
ds^2 = R^2 (-dt^2 \cosh^2 \rho +d \rho^2 + \sinh^2 \rho ~d\Omega^{2}_{3}).
\ee
A  folded closed string whose center lies at $\rho =0$  is spinning  at the equatorial plane
of $S^3$ and it is stretched from $\rho =0$ to $\rho =\rho_0$.
The polar angles are fixed, $\theta =\theta_1 = \pi/2$, the azimuthal
angles $\phi$ depends on time $t=\tau$ and $\rho$ is a function of $\sigma$
$$
\phi= \omega t, ~~~\rho = \rho(\sigma),
$$
so that the energy and the spin of the string are \cite{Gubser:2002tv}
\beqa\label{energy}
E =  {4R^2 \over 2\pi \alpha^{'}} \int^{\rho_0}_{0}
{ \cosh^2 {\rho}  d\rho  \over \sqrt{\cosh^2 {\rho}  - \omega^2 \sinh^2 {\rho} }}
\eeqa
\beqa\label{spin1}
S =  {4R^2 \over 2\pi \alpha^{'}} \int^{\rho_0}_{0}
{ \omega \sinh^2 {\rho} d\rho  \over \sqrt{\cosh^2 {\rho}  - \omega^2 \sinh^2 {\rho} }}
\eeqa
where $\rho_0$ is the maximal radial coordinate which depends on angular velocity
$$
\tanh^2 {\rho_0} = 1/\omega^2 .
$$
It follows from above formula that for large angular velocities $\omega \gg 1$
$$
\rho_0 \sim 1/\omega
$$
and the string is not stretched much compared to the radius of $AdS_5$, thus
it is a "short string". When $\omega$ approaches 1  from above $\omega  \sim 1$
$$
\rho_0 \sim {1\over 2} \ln {4 \over  1- 1/\omega^2 }
$$
the string is much longer than the radius of curvature of $AdS_5$ and
$\rho_0$ approaches the boundary of $AdS_5$ so that we have a "long string".

With the substitution $\omega \tanh \rho  = sin \varphi $
for the energy (\ref{energy}) and spin (\ref{spin1}) we shall get
\beqa\label{ }
E=  {4R^2 \over 2\pi \alpha^{'}\omega} \int^{\pi/2}_{0}
{   d\varphi  \over (1 - {1\over \omega^2 } \sin^2 {\varphi} )^{3/2}}
\eeqa
\beqa\label{ }
S =  {4R^2 \over 2\pi \alpha^{'} } [\int^{\pi/2}_{0}
{   d\varphi  \over (1 - {1\over \omega^2 } \sin^2 {\varphi} )^{3/2}}-
\int^{\pi/2}_{0}
{   d\varphi  \over (1 - {1\over \omega^2 }\sin^2{\varphi} )^{1/2}}
]
\eeqa
and using the identity
\beqa\label{ }
\int^{\pi/2}_{0}
{   d\varphi  \over (1 - k^2  \sin^2{\varphi})^{3/2}} =
{1 \over (1-k^2)} \int^{\pi/2}_{0}
(1 - k^2 \sin^2{\varphi} )^{1/2}    d\varphi
\eeqa
we can express the energy and the spin in terms of complete elliptic integrals
of the first $\textbf{ K}(k)$ and second $ \textbf{E}(k)$ kinds

\begin{eqnarray}\label{energyspin}
  ~ E  &=& {2   \sqrt{\lambda}   \over \pi  }   ~{1\over \omega }~
{1 \over (1- 1/\omega^2) }
{\textbf{ E} }(1/\omega) \\
~ S &=& { 2   \sqrt{\lambda}  \over\pi   }  ~   ( {1 \over 1- 1/\omega^2  }~
{ \textbf{E} }(1/\omega) - { \textbf{K} }(1/\omega)~) ,\label{energyspin1}
\end{eqnarray}
where we use the map
$
 R^2 = \sqrt{\lambda}  \alpha^{'}
$
and  that $k= 1/\omega$. These relations define  $ E(\lambda,\omega)$ and $ S(\lambda,\omega)$ in a
parametric form. Therefore in order to
express the energy as a function of spin $S$ one should invert the spin
function $\omega = \omega (\lambda, S)$. It is a nontrivial problem and to resolve  it we shall develop
a special analytical tools in the next sections.

The well known Legendre relation between complementary elliptic integrals has the
form
\be
\textbf{ E}(k) \textbf{K}(k^{'}) +  \textbf{K}(k)  \textbf{E}(k^{'}) -
\textbf{K}(k)  \textbf{K}(k^{'}) = {\pi \over 2}
\ee
where complementary module $k^{'}$ is defined as
$
k^{'2}   +  k^2   =1.
$
Using the above Legendre relation we shall get the duality relation
\begin{eqnarray}
{1\over \omega}E S^{'} + {1\over \omega^{'}}E^{'}S
- S  S^{'} =\frac{2\lambda}{\pi}
\end{eqnarray}
where
\be\label{dualityomega}
{1\over \omega^{2}} +{1\over \omega^{'2}}=1.
\ee
When $\omega \gg 1$ we have $\omega^{'} \sim 1$ and vice-versa, therefore it defines
exact map between energies and spins of {\it short and long strings}. Now it can be
rewritten in terms of anomalous dimension as
\begin{eqnarray}\label{duality}
{1\over \omega}~\gamma ~S{'} + {1\over \omega^{'}}~\gamma^{'} ~S
+({1\over \omega} + {1\over \omega^{'}} -1) S  S^{'} =\frac{2\lambda}{\pi}~.
\end{eqnarray}
This functional equation defines the duality map between anomalous
dimensions $\gamma=\gamma(\lambda,S)$  and $\gamma^{'}= \gamma(\lambda,S^{'})$
at complementary values of spins. Complementary values of spins are
defined through the equation (\ref{dualityomega}).
The question is: What is omega large and  omega small in terms of spins? In other
words which values of  S and $S^{'}$ are complementary? As we  shall see in the next section,
in the leading approximation  $1- {1\over \omega^{2}} =x =\frac{2 \sqrt{\lambda}}{\pi S}$ for large $S/\sqrt{\lambda}>>1$,
while $\frac{1}{\omega^2}=\frac{2 S}{\sqrt{\lambda}}$ for small $S/\sqrt{\lambda}<<1$.
Then equation (\ref{dualityomega}) gives
\be
\frac{S}{\sqrt{\lambda}}   \frac{S^{'}}{\sqrt{\lambda}}\approx \frac{1}{\pi}.
\ee
Thus the following regions are complementary-dual to  each other:
\be
 S \gg  \sqrt{\lambda}  \gg 1   ~~~~~~~~~~~\Leftrightarrow ~~~~~~~~~~~~~~~  \sqrt{\lambda}  \gg S   \gg 1 ,
\ee
that is the regions where the spin is much larger and much smaller than the coupling constant.

\section{\it Inverse Spin Function and Anomalous Dimension }

In order to study the behavior of the anomalous dimension in the subleading approximations
one should have well defined expansion of complete elliptic integrals at two
edges of the module space: $k = 1/\omega \sim 0 $ and $k = 1/\omega \sim 1$.
Using  formulas (\ref{ellipI}) and (\ref{ellipII})  we can represent the energy and  the
spin (\ref{energyspin}) in the following form (see Appendix for details)
\beqa\label{}
E = && {4R^2 \over 2\pi \alpha^{'}\omega}
\{~{1 \over (1- 1/\omega^2) } - {1\over 2\pi}
\sum^{\infty}_{n=0}  {\Gamma(n+1/2) \Gamma(n+3/2) \over n!(n+1)! }
 (1-1/\omega^2)^{n}   \times\nn\\
 &\times& (\ln(1-1/\omega^2) + \psi(n+ 1/2)  +\psi(n+3/2) - \psi(n+1) - \psi(n+2)) ~ \},
\eeqa
\begin{eqnarray}
S   &=& {4R^2 \over 2\pi \alpha^{'}}
\{~{1 \over (1- 1/\omega^2) } - {1\over 2\pi}
\sum^{\infty}_{n=0}  {\Gamma(n+1/2) \Gamma(n+3/2) \over n!(n+1)! }
 (1-1/\omega^2)^{n}   \times\nn\\
 &\times& (\ln(1-1/\omega^2) + \psi(n+ 1/2)  +\psi(n+3/2) - \psi(n+1) - \psi(n+2))  +\nn\\
 &+& {1\over 2\pi}
\sum^{\infty}_{n=0}  ({ \Gamma(n+1/2) \over n!})^2
 (1-1/\omega^2)^n    \times\nn\\
 &\times& (\ln(1-1/\omega^2) + 2  \psi(n+ 1/2) - 2 \psi(n+1))~
  \}.
\end{eqnarray}
The explicit expansion of the first few terms has the form
\beqa\label{}
E(\omega)=  {4R^2 \over 2\pi \alpha^{'}\omega}
\{~{1 \over (1- 1/\omega^2) } - {1\over 4}[\ln(1-1/\omega^2)- 4 \ln 2 +1]-\nn\\
-{3\over 32}(1-1/\omega^2)[\ln(1-1/\omega^2)-4 \ln2 +{13\over 6}]+...\},
\eeqa
\begin{eqnarray}
S(\omega)  = {4R^2 \over 2\pi \alpha^{'}}
\{~{1 \over (1- 1/\omega^2) } + {1\over 4}[\ln(1-1/\omega^2)-  4 \ln 2 -1]+\nn\\~
+{1\over 32}(1-1/\omega^2)[\ln(1-1/\omega^2)-4 \ln2 +{3\over  2}]+...  \}.
\end{eqnarray}
We can represent now the energy and  the
spin (\ref{energyspin}) in the form convenient for expansion. The spin can be represented in the form
\begin{eqnarray}\label{spin}
\CS ={ \pi  \over 2   } { S   \over  \lambda^{1/2} } =
\{~{1 \over x } +\sum^{\infty}_{n=0}x^n (c_n \ln x + b_n) \}
\end{eqnarray}
and the energy as
\be\label{EEnergy}
 \CE     = { \pi  \over 2   } { E   \over  \lambda^{1/2} }  =
\sqrt{1-x}~(~{1 \over x} +\sum^{\infty}_{n=0}x^n(d_n \ln x +h_n )~)
\ee
where $x=1-1/\omega^2$ and
\beqa
d_n &=&-{1\over 2^{2n+2}}{(2n-1)!!(2n+1)!! \over n!(n+1)! },~~~~ \\
h_n &=& d_n~
[~\sum_{k=1}^{n} {2 \over  k (2k-1)} + {1 \over (n+1)(2n+1)}-4 \ln 2 ]\nn\\
c_n &=&d_n +{1\over 2^{2n+1}}({(2n-1)!! \over n!  })^2,~~~~ \nn\\
b_n &=& h_n~ + ~{1\over 2^{2n+1}}({(2n-1)!! \over n!  })^2
[~\sum_{k=1}^{n} {2 \over  k (2k-1)} -4 \ln 2 ] . \nn
\eeqa
The explicit expression for the first few coefficients is
\begin{eqnarray}
 d_0 = -{1 \over 4}, ~~&d_1 = -{3 \over 32},~~&d_2 =-{15 \over 2 \cdot 128}, \nn\\
 h_0= \ln 2 - {1 \over 4},~~ &h_1 = {3 \over 8} \ln 2 - {13 \over 64},~~
&h_2 ={15 \over 64} \ln 2 - {9 \over 64}, \nn\\
 c_0 = {1 \over 4}, ~~&c_1= {1 \over 32},~~&c_2 ={3 \over 2 \cdot 128}, \nn\\
 b_0= -\ln 2 - {1 \over 4},~~&b_1 = -{1 \over 8} \ln 2 + {3 \over 64},~~
&b_2 =-{6 \over 128} \ln 2 + {3 \over 128}.
\end{eqnarray}

\begin{figure}
\centerline{\hbox{\psfig{figure=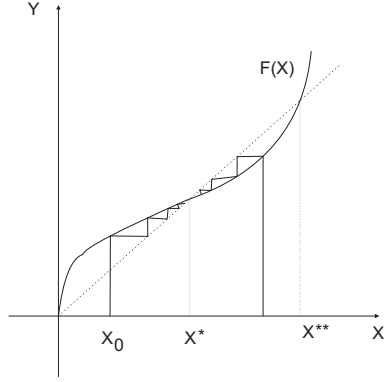,width=5cm}}}
\caption[fig]{ The iteration function $F(x) = 1/(\CS- c_0 \ln x - b_0)$
has two nontrivial fixed points $x^*$ and $x^{**}$,
which are the solutions of the equation $F(x )=x $ . The iteration is defined as
$x_n = FF....F(x_0)$ and converges to the stable fixed point $x_n \rightarrow x^*$.
In the given case the
fixed point $x^*$ is a stable one, while the second one $x^{**}$ is unstable and
unphysical.}
\label{fig1}
\end{figure}
In order to invert the spin function $x=x(\CS)$ we shall define
 the function $x^*(\CS)$
as a solution of the "reduced" spin equation  (\ref{spin})
\be\label{fixetpointeq}
\CS =
 {1 \over x^* } + c_0 \ln x^* + b_0 .
\ee
The benefit of using the function  $x^*(\CS)$ is that the inverse spin function $x=x(\CS)$
can be expressed in terms of  $x^*(\CS)$.
The function $x^*(\CS)$ can be found by iteration of the
following map (see Fig 1)
\be
F(x)={1 \over \CS - c_0 \ln x - b_0}.
\ee
Indeed the iteration $x_n = FF....F(x_0)$ which starts from $x_0 = {1 \over \CS}$ gives
\be
x_0 = {1 \over \CS} \rightarrow{1 \over \CS - c_0 \ln 1/\CS - b_0}\rightarrow
{1 \over \CS -c_0 \ln {1 \over \CS - c_0 \ln 1/\CS - b_0}   - b_0} \rightarrow ...
\rightarrow x^*(\CS).
\ee
and has the property that its fixed point $F(x^*)=x^*$ is a solution of the
equation (\ref{fixetpointeq}). The iteration process can be represented in the
form of infinite product
\beqa\label{infiniteproduct}
x_0 &=&{1 \over \CS}\nn\\
x_1&=&{1 \over \CS}\cdot {1 \over 1- A/\CS} ~~~~\nn\\
x_2&=&{1 \over \CS}\cdot {1 \over 1- A/\CS}  \cdot
{1 \over 1 +c_0{ 1\over \CS}  {1 \over  (1- A/\CS)}\ln (1- A/\CS)} \\
&&..............................................\nn\\
x^*&=&{1 \over \CS}\cdot {1 \over 1- A/\CS}  \cdot
{1 \over 1 +c_0{1 \over \CS}   {\ln (1- A/\CS)\over  (1- A/\CS)}} \cdot
{1 \over 1 +c_0 {1\over \CS}   {1 \over 1- A/\CS}   {1 \over  (1 +{c_0 \over \CS}
  {\ln (1- A/\CS)\over  (1- A/\CS)})}\ln (1 +{c_0 \over \CS}
 {\ln (1- A/\CS)\over  (1- A/\CS)})} \cdot \cdot ~\cdot
\nn
\eeqa
where
\be
A= c_0 \ln(1/\CS) +b_0= -{1\over 4}(\ln 16 \CS +1).
\ee
It follows therefore that the fixed point solution $x^*(\CS)$ can be represented in the
form of infinite product in which $x_1/x_0$   contains the
leading terms  ${ \ln^n \CS  \over \CS^{n} }$, the $x_2/x_1$ contains
subleading terms ${ \ln^n \CS  \over \CS^{n+1} }$, the $x_3/x_2$ contains
next to subleading terms ${ \ln^n \CS  \over \CS^{n+2} }$ and so on.
In general, each subsequent $k$-th term in the infinite product gives
the higher orders terms in the expansion over ${ \ln^n \CS  \over \CS^{n+k} }$.
The above infinite product representation (\ref{infiniteproduct}) of the
fixed point solution $x^*(\CS)$ allows also
the representation in the form of infinite sum
\be\label{infinitesum}
x^*(\CS)={1 \over \CS} \cdot [1+X_1\left({A \over \CS}\right)]  \cdot
[1+ {1 \over \CS } X_2({A \over \CS})] \cdot ....={1 + X_1\over \CS} +
{ (1+X_1 ) X_2  \over \CS^2} + ...
\ee
The infinite product representation (\ref{infiniteproduct}) can be conveniently used
when one should compute the logarithm of  $x^*$ and the infinite sum representation
(\ref{infinitesum}) when one should compute polynomials of  $x^*$.

The advantage of using the fixed point solution of the reduced spin equation
is that we can find the inverse spin function $x=x(\CS)$ in any
required approximation using the fixed point solution $x^*(\CS)$.
The inverse spin function  $x(\CS)$
can be represented as a sum of the fixed point solution $x^*(\CS)$ plus terms
which are next to subleading and higher
\be\label{nexttonext}
x(\CS) = x^*(\CS) + \sum^{\infty}_{n=0} a_n {(\ln 1/\CS)^n \over \CS^{n+2}} +...= x^*  +
\delta x_2 +...
\ee
Substituting the expansion (\ref{nexttonext}) into (\ref{spin}) we shall get
the next to subleading term $\delta_2 x$ as a function
of $x^*$
\be\label{subleading}
  \delta x_2  =
c_1 (x^*)^3 \ln x^* .
\ee
In order to find leading and subleading terms in the anomalous dimension
we have to keep quadratic in $x$ terms in the
$\CE - \CS$ and use the
relations which follow from (\ref{nexttonext}) and (\ref{subleading}).
Indeed the next to subleading term $\delta x_2$ in $x(S)$ will generate the following subleading
terms when multiplied by the logarithm of $x^*$ or divided by $x^*$
\be
\delta x_2 ~\ln x^* ~~~\rightarrow ~~~~
\sum^{\infty}_{n=0} a_n {(\ln 1/\CS)^{n+1} \over \CS^{n+2}},~~~~~ {\delta x_2 \over x^*} ~~~
\rightarrow ~~~
\sum^{\infty}_{n=0} a_n {(\ln 1/\CS)^{n } \over \CS^{n+1}}
\ee
therefore we shall get
\beqa
 \CE -\CS
=(d_0-c_0)\ln x + (h_0 - b_0  -{1 \over 2} )+
(d_1  -  c_1- {1 \over 2} d_0 )x  \ln x + (h_1  -b_1- {1 \over 2} h_0 -{1 \over 8} ) x+\nn \\
+(d_2 -  c_2 - {1 \over 2} d_1- {1 \over 8} d_0 )x^2  \ln x +(h_2 -b_2
- {1 \over 2} h_1  -{1 \over 8} h_0 -  {1 \over 16} ) x^2 =
\nn\\
(d_0-c_0)\ln x^* + (d_0-c_0) {\delta_2 x \over x^*} + (h_0 - b_0  -{1 \over 2} )+
(h_1  -b_1- {1 \over 2} h_0 -{1 \over 8} ) x^* +
\nn\\
+(d_2 -  c_2 - {1 \over 2} d_1- {1 \over 8} d_0 ) (x^*)^2  \ln x^* +(h_2 -b_2
- {1 \over 2} h_1  -{1 \over 8} h_0 -  {1 \over 16} ) (x^*)^2 =
\nn\\
=(d_0-c_0)\ln 1/\CS + (h_0 - b_0  -{1 \over 2}) + (d_0-c_0)\sum^{\infty}_{n=1}
{(c_0 \ln 1/\CS  +b_0 )^n \over n \CS^n } +
\nn\\
+(d_0-c_0)c_0 \sum^{\infty}_{n=1} \sum^{\infty}_{k=1} {1 \over n \CS}
\left({c_0 \ln 1/\CS   \over   \CS  }\right)^{n+k-1}+
\nn\\
+(d_0-c_0)c_1 \sum^{\infty}_{n=1}n { (c_0\ln 1/\CS +b_0)^{n-1} \ln 1/\CS \over \CS^{n+1} }
\nn\\
+ (h_1  -b_1- {1 \over 2} h_0 -{1 \over 8} ) {1 \over   \CS }
+ (h_1  -b_1- {1 \over 2} h_0 -{1 \over 8} )\sum^{\infty}_{n=1}
{(c_0 \ln 1/\CS + b_0)^n \over  \CS^{n+1} }+
\nn\\
+(d_2 -  c_2 - {1 \over 2} d_1- {1 \over 8} d_0 )
\sum^{\infty}_{n=1}n{ (c_0\ln 1/\CS +b_0)^{n-1} \ln 1/\CS \over \CS^{n+1} }+
\nn\\
+(h_2 -b_2
- {1 \over 2} h_1  -{1 \over 8} h_0 -  {1 \over 16} ) {1\over \CS^2} +...
\CO \left({ \ln^m 1/\CS  \over \CS^{m+2} }\right)~~.~~~~~~~~~
\eeqa
Representing this expansion in the following from
\be
\CE -\CS = {1 \over 2} \ln \CS + \rho_0  +  \sum^{\infty}_{n=1} \rho_{nn}{ \ln^n  \CS  \over  \CS^n }
+  \sum^{\infty}_{n=1} \rho_{nn-1}{ \ln^n  \CS  \over  \CS^{n+1}  }
+ \rho_1 {1 \over \CS} + \rho_2 {1 \over \CS^2}
\ee
we can derive all these coefficients
\beqa
\rho_0 &=& h_0 - b_0  - {1 \over 2}  = 2 \ln 2 - {1 \over 2}
\nn\\
\rho_{nn} &=&{(-1)^n \over n} (d_0-c_0) c^{n}_{0} =
{1 \over 2}{ (-1)^{n+1}   \over 4^n } {1\over n}
\nn\\
\rho_{nn-1} &=&   (-1)^{n}  c^n_0    ~[~(d_0-c_0) b_0
+(d_0-c_0) c_0 \sum^{n}_{k=1}{1 \over k}
+ h_1 -b_1 - {1 \over 2} h_0  - {1 \over 8}~]~
  +\nn\\
&&+(-1)^{n} n~ c^{n-1}_0 ~[~(d_0-c_0) c_1  + d_2 -c_2 - {1 \over 2} d_1  - {1 \over 8}d_0~]~
\nn\\
&=& {1 \over 2}{ (-1)^{n+1}   \over  4^n }~({n +4 \over 16}
+ {1 \over 4} \sum^{n}_{k=1}{1 \over k}- \ln 2 )
\nn\\
\rho_1 &=& h_1  -b_1- {1 \over 2} h_0 -{1 \over 8}+ (d_0-c_0) b_0
= {1 \over 2} \ln 2 - {1 \over 8}
\nn\\
\rho_2 &=& (d_0-c_0)(c_0 b_0 +b^{2}_{0}/2) +
(h_1  -b_1- {1 \over 2} h_0 -{1 \over 8})b_0 + \nn\\
&+&h_2 -b_2 - {1 \over 2} h_1  -{1 \over 8} h_0 -  {1 \over 16}=
-{1 \over 4} \ln^2 2 +{7 \over 32} \ln 2 - {1 \over 64}.
\eeqa
Finally the dependence of the anomalous dimension from the spin will take the form
\be\label{E-S}
E-S = {\sqrt{\lambda} \over \pi} \ln  (S/\sqrt{\lambda}) + f_0
+  \sum^{\infty}_{n=1} f_{nn}{ \ln^n (S/\sqrt{\lambda}) \over   S^n }
+  \sum^{\infty}_{n=1} f_{nn-1}{ \ln^n   (S/\sqrt{\lambda})  \over   S^{n+1}  }
+ f_1 {1 \over  S} + f_2 {1 \over  S^2}+...
\ee
where
\beqa\label{fs}
f&=&{\sqrt{\lambda} \over \pi}\nn\\
f_0 &=& {\sqrt{\lambda} \over \pi}~ (2 \rho_0 +  \ln (\pi /2))=
{\sqrt{\lambda} \over \pi}  (\ln 8 \pi -1)
\nn\\
f_1 &=& {4 \lambda \over \pi^2} ~(\rho_1  + ~\rho_{11} \ln (\pi/2) )
= { \lambda  \over 2\pi^2}  (\ln 8 \pi -1)
\nn\\
f_2 &=&{8 \lambda^{3/2} \over \pi^3} ~(\rho_2 + \rho_{10}\ln (\pi/2) +\rho_{22}\ln^2 (\pi/2))=\nn\\
&=& { \lambda^{3/2}  \over 8 \pi^3 }  (- \ln^2 8\pi +{9 \over 2} \ln 8\pi -4\ln 2-1 )
\nn\\
f_{nn} &=&({2 \lambda^{1/2} \over \pi })^{n+1} ~  \rho_{nn}= (-1)^{n+1}
{\lambda^{{ n+1\over 2}} \over (2\pi)^{n+1}}{ 2  \over n   }=\frac{(-1)^{n+1}}{2^n n}f^{n+1} ,~~~~n=1,2,...
\nn\\
f_{nn-1} &=&  ({2 \lambda^{1/2} \over \pi })^{n+2} ~
(~\rho_{nn-1} + \rho_{n+1n+1} (n+1)\ln (\pi/2)~)=\nn\\
&=&  (-1)^{n+1}
{\lambda^{{ n+2 \over 2}} \over (2\pi)^{n+2}}~
(~ {n+4 \over 2} + 2 \sum^{n}_{k=1}{1\over k} -  2 \ln 8\pi  ~),~~~~n=1,2,...
\eeqa
and
\be
\sqrt{\lambda}  \gg 1 ,~~~~ {S \over \sqrt{\lambda}} \gg 1 .
\ee
We close this Section by noticing that the iteration procedure described above can be
generalised to give us all the terms in the large $S$ expansion of the strong coupling anomalous dimension.
For example, if one wanted to compute the infinite series of the coefficients $f_{nn-2}$ in front of the
sub-subleading terms $\frac{\ln^n S}{S^{n+2}}$
in \eqref{E-S} one should define an $x^*(\CS)$ which is the solution of the equation
\be\label{fixetpointeq2}
\CS =
 {1 \over x^* } + c_0 \ln x^* + b_0 +x^*(c_1 \ln x^* + b_1).
\ee
Then the inverse spin function  $x(\CS)$
can be represented as a sum of the fixed point solution $x^*(\CS)$ plus terms
which are next to sub-subleading and higher
\be\label{nexttonextt0next}
x(\CS) = x^*(\CS) + \sum^{\infty}_{n=0} t_n {(\ln 1/\CS)^n \over \CS^{n+3}} +...= x^*  +
\delta x_3 +...
\ee
$\delta x_3$ and as a result the coefficients $t_n$ can be determined
by demanding that \eqref{nexttonextt0next} satisfies \eqref{spin}
up to subleading order, i.e. $\frac{\ln^n S}{S^{n+1}}$.
More precisely one gets,
\be\label{d3x}
\delta x_3=c_2 \ln x^* \, \, x^{*4},
\ee
where $x^*$ is the solution of \eqref{fixetpointeq2}.
Then one can substitute the solution \eqref{nexttonextt0next} in the
expression for the energy minus spin, $E(x)-S(x)$, to get the anomalous dimension up to sub-subleading order, that is up
to order $\frac{\ln^n S}{S^{n+2}}$.

Let us only make two observations.
The first is that since we are interested in the sub-subleading terms we should keep
terms which mostly go as $x^3 (\ln x)$ or $x^3$ in the expansion for ${\cal E}-{\cal S}$
(see equations \eqref{spin}, \eqref{EEnergy}).
The second is that since $x^*$ is found by iteration of the following map
\be
F(x)={1 \over \CS - c_0 \ln x - b_0-x(c_1 \ln x + b_1)}
\ee
one should make as many iterations needed for reaching the sub-subleading order in the expression for $E-S$.

\section{\it Functional and Reciprocity  Relation}
As we mentioned in the Introduction, the operators dual to the string states we are considering are of the form
\be
{\cal O}_s=tr(Z D^S_{+}Z)+...
\ee
and as such they belong to a $SL(2,R)$ subgroup of the full $PSU(2,2/4)$ group of $N=4$ SYM.
They are labelled by the conformal spin $s=\frac{1}{2}(S+\Delta)$.
Thus, it is natural to assume that the anomalous dimension $\gamma$ of these operators depends on $S$
only through the conformal spin $s=1+S+\frac{1}{2}\gamma$. This assumption leads to the following functional relation
for $\gamma$ \cite{Basso:2006nk,Dokshitzer:2006nm}
\beqa\label{functionalg}
\gamma(S)=P(S+\frac{1}{2}\gamma(S)).
\eeqa

\eqref{functionalg} can be inverted to give us a functional relation for $P$. Namely,
\beqa\label{functionalP}
P(S)=\gamma(S-\frac{1}{2}P(S)).
\eeqa
Assuming that the anomalous dimension $\gamma$ is known up to some order one can determine
the function $P$ and vice-versa. This can be achieved using the following relation
\beqa\label{Pderivatives}
P(S)=\sum_{k=1}^{\infty} \frac{1}{k!}(-\frac{1}{2}\partial_S)^{k-1}\gamma^k(S).
\eeqa
Using the infinite series for $f_{nn}$ given in \eqref{fs}
we see that the coefficients of the leading logarithms $\frac{\ln^nS}{S^n}$ in the expansion of $\gamma(S)$ are fully determined
by the scaling function $f$. This is consistent with the functional relation \eqref{functionalg} and one can rewrite $\gamma(S)$
as
\beqa\label{anomalous}
\gamma(S)={\sqrt{\lambda} \over \pi} \ln  (S/\sqrt{\lambda})
+  \sum^{\infty}_{n=1} \frac{(-1)^{n+1}}{2^n n}f^{n+1}{ \ln^n (S/\sqrt{\lambda}) \over   S^n }+...=f\ln(S+\frac{1}{2}f \ln S+...)+...
\nonumber\\
\eeqa
From \eqref{anomalous} it is obvious that the functional relation \eqref{functionalg} holds for the leading logarithmic terms
of the form $\frac{\ln^nS}{S^n}$, with $P$ given by
\be
P(S)=f \ln S+... ,
\ee
where the dots stand for subleading terms in $S$.

Another important consequence of the form of $f_{nn}$'s is that the function $P$ obeys the so called ''simplicity'' condition
\footnote{We should note that this ''simplicity'' condition is broken in perturbation theory for twist two operators \cite{Beccaria:2009vt}
and twist three
operators \cite{Beccaria:2009eq} at critical wrapping order. We thank Valentina Forini for pointing this out.}
\cite{Beccaria:2010tb}, which states that $P$ is simpler than $\gamma$ in the sense that it contains no leading logarithms in its expansion
\be\label{Pexpansion}
P(S) = {\sqrt{\lambda} \over \pi} \ln  (S/\sqrt{\lambda}) + {\hat P}_0
+  \sum^{\infty}_{n=1} {\hat P}_{nn-1}{ \ln^n   (S/\sqrt{\lambda})  \over   S^{n+1}  }+...
\ee
Should $P$ contained those terms the value of the coefficients $f_{nn}$ would be different from that of
\eqref{fs}.

Furthermore, it is conjectured
 that $P(S)$ satisfies a ''parity preserving'' or ''reciprocity'' relation.
 This kind of relation was first observed in the context of deep inelastic scattering (DIS) in QCD \cite{Lipatov1,Lipatov2,Moch:2004pa,Vogt:2004mw}.
 It is a relation involving the splitting functions $P_s=P_t=P(x)$
 \be
P(x)=-x P\Big(\frac{1}{x}\Big).
\ee
These splitting functions can be related to the anomalous dimension of twist two operators
 through a Mellin transform.
 In the case of the maximally supersymmetric theory
 it can be cast in the form \cite{Basso:2006nk}
\beqa\label{reciprocity}
P(S)=\sum_{k=0}^{\infty}\frac{c_k(\ln C)}{C^{2k}}.
\eeqa
In \eqref{reciprocity}, $C$ is the ''bare'' quadratic Casimir operator of the $SL(2,R)$ group given by
$C=s_0(s_0-1)$, where $s_0=\frac{1}{2}(S+\Delta_0)=S+1$ is the value of the conformal spin at the classical level.
As a result, $C=S(S+1)$.
Equation \eqref{reciprocity} implies relations between the subleading terms in the expansion \eqref{E-S}.
These relations can be obtained by comparing the coefficients of the subleading logs in equation \eqref{functionalP}.
More precisely, by equating \eqref{Pexpansion} and \eqref{reciprocity} it is immediate to get that $P_{2n\,2n-1}=0$.
Then, one can exploit \eqref{Pderivatives} and compare the coefficients of the subleading logs to an even power,
i.e. $\frac{\ln^{2n} S }{S^{2n+1}}$ to get
the aforementioned relations between the subleading terms in the expansion \eqref{E-S}.
The comparison of the odd terms $\frac{\ln^{2n+1} S }{S^{2n+2}}$ determine the value of the non-zero coefficients
${\hat P}_{2n+1\,2n}$ of the subleading terms
appearing in the expansion of $P(S)$ \eqref{Pexpansion}.

The reciprocity relation \eqref{reciprocity} has been verified to hold in perturbation theory up
to five loops for the anomalous dimension of twist two operators in $N=4$ SYM (see \cite{Beccaria:2010tb} and references therein).
At strong coupling the first two MVV relations
\beqa\label{recTseytlin}
f_{1}&=& \frac{1}{2}f (f_0 +1) \nonumber \\
f_{21}& =& \frac{1}{16}f [f^3 - 2f^2(f_0 +1) - 16f_{10}]\nonumber \\
&.&\nonumber \\
&.&\nonumber \\
&.&\nonumber \\
\eeqa
were verified to hold \cite{Beccaria:2008tg}.
Since now we have the infinite series of the coefficients of the subleading logs in the strong coupling expansion \eqref{E-S}
it is possible to check if the relations following from the reciprocity property of the function $P$ hold at any level.
Here we focus on the next relation in the sequence of \eqref{recTseytlin}.
As mentioned above, this can be derived by equating the coefficients multiplying the
$\frac{\ln^{4} S }{S^{5}}$ term in \eqref{reciprocity}.
To this end,we have to decide how many terms we should keep on the right hand side of \eqref{reciprocity}.
By taking 5 derivatives of a generic term $\frac{\ln^a S}{s^b}$ appearing in the right hand side of \eqref{reciprocity} we
take terms like $\frac{\ln^{a-i} S}{S^{b+5}}$, where $i=0,1,...,5$. For these terms to be equal to $\frac{\ln^{4} S }{S^{5}}$
we have to demand that $b=0$ and $a=9,8,7,6,5,4$. Notice that if we differentiate once more, take 6 derivatives of $\frac{\ln^a S}{S^b}$
then the denominator of the result would be $S^{b+6}$ which implies that $b=-1$. But this is impossible since only negative powers of $S$
appear in the expansion of the anomalous dimension or any positive power of it. As a result the last term which can be contribute
to the coefficient of $\frac{\ln^4 S}{S^5}$
in the right hand side of \eqref{reciprocity} is the $5^{th}$ derivative of $\gamma^6(S)$.

By isolating the appropriate terms we get the next relation in the sequence \eqref{recTseytlin}. This reads
\beqa\label{recus}
f_{43}+[f^2_{22}+2f_{11}f_{33}+2 f f_{32}+ f_{44}(2 f_0-\frac{5}{2}f)] +f[3f_{11}f_{22}+\frac{3}{2}f f_{21}+f_{33}(-\frac{35}{8}f+ 3f_0)]
\nonumber \\
+f^2[\frac{3}{4}f^2_{11}+\frac{1}{2}f f_{10}+f_{22}
(-\frac{65}{24}f+\frac{3}{2}f_0)]+f^3[\frac{1}{16}f f_1+f_{11}
(-\frac{125}{12 \,\,16}f+\frac{1}{4}f_0)]~~~~~~~~~~
\nonumber \\
+f^5[\frac{25}{12 \,\,32}f-\frac{1}{32}f_0] =0.~~~~~~~~~~~~~~~~~~~~~~~~~~~~~~~~~~~~~~~~~~~
\eeqa
Now we can use \eqref{fs} to find and substitute the values for the various coefficients needed in \eqref{recus} to find that
this equation is, indeed, satisfied.
We conclude that all the constraints following from the reciprocity relation \eqref{reciprocity} are likely to be satisfied
in string perturbation theory too.

\section{Acknowledgements}
We would like to thank E. Floratos for stimulating discussions
and Valentina Forini for useful comments on the first version of this work.

\section{\it Appendix}
For the small values of the module $\vert k^2 \vert < 1$ the elliptic
integral of the first kind  $\textbf{ K}(k)$ has the well known expansion
\beqa
\textbf{ K}(k)={\pi \over 2}\{1 +
\sum^{\infty}_{n=1}[{(2n-1)!! \over 2n!!}]^2 \cdot  k^{2n}  ~~\},~~~~~~
\vert k^2 \vert < 1  \nn
\eeqa
and for the large values $\vert 1-k^2 \vert < 1 $ we have found the following expansion
\begin{eqnarray}\label{ellipI}
\textbf{K}(k)  = - {1\over 2\pi}
\sum^{\infty}_{n=0}  ({ \Gamma(n+1/2) \over n!})^2
 (1-k^2)^n    \cdot
 [~\ln(1-k^2) + 2  \psi(n+ 1/2) - 2 \psi(n+1)~],
\end{eqnarray}
where $\psi(z)$ is the digamma function.
For the elliptic integral of the second kind $\textbf{E}(k)$
the expansion for small values of the module $\vert k^2 \vert < 1$  is
\beqa
\textbf{E}(k)={\pi \over 2}\{1 -
\sum^{\infty}_{n=1}[{(2n-1)!! \over 2n!!}]^2 \cdot {k^{2n}\over 2n-1} \},~~~~~~
\vert k^2 \vert < 1 \nn
\eeqa
and for the large values $\vert 1-k^2 \vert < 1$ it is
\begin{eqnarray}\label{ellipII}
\textbf{E}(k) &=&1 - {(1-k^2)\over 2\pi}
\sum^{\infty}_{n=0} {\Gamma(n+1/2) \Gamma(n+3/2) \over n!(n+1)! }
(1-k^2)^n    \cdot\nn\\
 &\cdot& [~\ln(1-k^2) + \psi(n+ 1/2) +\psi(n+3/2) - \psi(n+1) - \psi(n+2) ~].
\end{eqnarray}

\vfill
\end{document}